# Safeguarding Quantum Key Distribution through Detection Randomization

Thiago Ferreira da Silva, Gustavo C. do Amaral, Guilherme B. Xavier,
Guilherme P. Temporão, and Jean Pierre von der Weid

*Abstract*—We propose and experimentally demonstrate a scheme to render the detection apparatus of a Quantum Key Distribution system immune to the main classes of hacking attacks in which the eavesdropper explores the back-door opened by the single-photon detectors. The countermeasure is based on the creation of modes that are not deterministically accessible to the eavesdropper. We experimentally show that the use of beamsplitters and extra single-photon detectors at the receiver station passively creates randomized spatial modes that erase any knowledge the eavesdropper might have gained when using bright-light faked states. Additionally, we experimentally show a detector-scrambling approach where the random selection of the detector used for each measurement – equivalent to an active spatial mode randomization – hashes out the side-channel open by the detection efficiency mismatch-based attacks. The proposed combined countermeasure represents a practical and readily implementable solution against the main classes of quantum hacking attacks aimed on the single-photon detector so far, without intervening on the inner working of the devices.

## I. INTRODUCTION

QUANTUM key distribution (QKD) [1] benefits from the laws of quantum physics to provide absolutely secure communication [2] between two communicating parties (Alice and Bob), even if imperfect devices are used [3-5]. Imperfections on the equipment used in a QKD system can be related to sources that emit multi-photon pulses which enable the photon-number splitting attack [1]. Clever solutions based on fundamental principles were used to overcome such imperfections, as in the decoy states method [6-9], which enable a more efficient use of imperfect photon sources – more specifically weak coherent states with moderate multi-photon probability – for QKD.

Recently a critical point was recognized: back-doors may be open in some physical devices comprising the QKD system, specially the single-photon detector (SPD). The flaws may be explored by an eavesdropper (Eve) for side-channel attacks [10-23], which can jeopardize the security of the protocol. These quantum hacking attacks are interventions caused by Eve from the outside of Bob's station by high-jacking the detection apparatus – whose response can be predicted in some degree or even manipulated. In all cases, the attacks make it possible for an eavesdropper to gain information without being noticed, i.e., achieving a critically high mutual information with Alice and Bob without exceeding the upper threshold of the quantum bit error rate (QBER).

The hacking schemes basically aim on two key points: exploring the imperfect nature of the SPD – efficiency mismatched-based attacks [10-12] – or externally forcing a deterministic result on the detection equipment – bright-light-based attacks [14-19]. Different countermeasures to avoid detector-aimed quantum hacking attacks have been presented [10-28]. Despite being effective for the proposed specific end, i.e., closing a specific back-door, the solutions have no guarantees of being final, in the sense that the vulnerabilities depend on the physical implementation of the devices and the deployment of the systems. The counter-measures give, in the best case, *ad hoc* protection over some class of attacks.

Measurement-device-independent QKD has been recently proposed to eliminate detection [29] and/or state preparation [30] loopholes. Although making use of sophisticated schemes based on long-distance interference at a third remote station [31] and (reversely or not) entangled systems, practical implementations of the schemes have been reported [32-35]. Nevertheless, the well-established traditional BB84-like QKD systems still lack a definitive solution against quantum hacking.

We propose a practical solution that extends over a broad range of known classes of quantum hacking attacks aimed at the detection equipment. The scheme is based on fundamental randomization of input modes to the detection apparatus inside Bob's station, thus not deterministically accessible to the eavesdropper. Here we show that the use of beamsplitters and extra detectors at Bob's station renders its apparatus immune to bright-light based attacks, as the blinding- and faked-states attacks [20,21]. The eavesdropper can no longer manipulate

Manuscript received August 1, 2014. This work was supported by the Brazilian agency FAPERJ and CNPq. G.B.X. acknowledges support of CONICYT PFB08-024, Milenio P10-030-F and FONDECYT no. 11110115.

T. Ferreira da Silva is with the Center for Telecommunication Studies, Pontifical Catholic University of Rio de Janeiro, Rio de Janeiro, RJ, Brazil. He is also with the Optical Metrology Division, National Institute of Metrology, Quality and Technology, Duque de Caxias, RJ, Brazil (e-mail: thiago@opto.cetuc.puc-rio.br).

G. B. Xavier is with the Departamento de Ingeniería Eléctrica, Universidad de Concepción, Concepción, Chile. He is also with the Centre for Optics and Photonics and with the MSI-Nucleus on Advanced Optics, Universidad de Concepción, Concepción, Chile (e-mail: gxavier@udec.cl).

G. C. do Amaral, G. P. Temporão and J. P. von der Weid are with the Center for Telecommunication Studies, Pontifical Catholic University of Rio de Janeiro, Rio de Janeiro, RJ, Brazil (e-mail: gustavo@opto.cetuc.puc-rio.br; temporao@opto.cetuc.puc-rio.br; vdweid@opto.cetuc.puc-rio.br).

the detectors without leaving a strong signature which is monitored by the counting statistics of the detectors. Correlation between detectors in equivalent spatial modes reveals the attack, without intervening on the inner workings of the devices.

Additionally, we experimentally show that, as suggested in [12], a detector-scrambling strategy employed by Bob dynamically alters the detector used for measurement under the chosen basis, counteracting the attacks based on the detection efficiency mismatching, like the time-shift attack [12-13]. This proposal is readily implementable through the random application of a random rotation to invert the set of mutually unbiased detection bases. This is equivalent to an active spatial mode randomization since, after the basis choice, a random SPD will effectively be used for the detection of a particular state.

The assumptions for the proposed counter-measure are that the eavesdropper has no access on the inner working of the devices and cannot deterministically manipulate the splitting ratio of the BS. The main drawbacks of our proposed scheme against quantum hacking are the increase in the number of detectors when compared to the traditional setup for the BB84 protocol and an increase in the dark count rate. Nevertheless, the scheme provides a practical readily available real-world solution against, at least, those known classes of quantum hacking attacks.

## II. QUANTUM HACKING AIMED ON THE SPD

### A. The BB84 Context

In the BB84 protocol [1], Alice prepares each qubit in a single-photon state, according to a random choice between four states that form two pairs of orthogonal states in canonically conjugated bases in a bi-dimensional Hilbert space. Considering polarization states encoding, the bases may be sorted from rectilinear (⊕), composed by horizontal and vertical states of polarization (SOPs); or diagonal (⊗), composed by diagonal (+45°) and anti-diagonal (-45°) states-of-polarization (SOPs).

Bob randomly chooses the measurement basis for each incoming qubit and has a deterministic or probabilistic result according to the overlap between his own and Alice's chosen bases. Bob can choose to measure on ⊕ or ⊗ bases by turning the half-wave plate (HWP) to 0 or 22.5°, respectively.

Bob's typical detection apparatus for polarization qubits includes a HWP to change the measurement basis, a polarizing beamsplitter (PBS) and two SPDs, as seen in Fig. 1a.

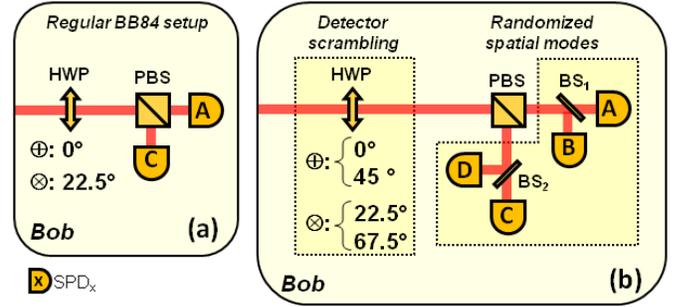

Fig. 1. (a) Typical detection apparatus on Bob's station for BB84-based QKD protocol. (b) Proposed countermeasure scheme with passive randomized spatial modes – employing two BSs and four SPDs – and active detector scrambling – increasing the set of angles at the HWP.

If Alice's and Bob's bases match, the photon is routed to a deterministic spatial mode and is delivered to the SPD corresponding to the transmitted qubit. Otherwise, it becomes a superposition of the PBSs eigenstates and the photon randomly emerges at one output spatial mode, which can then be detected with an SPD. During the basis reconciliation in the post-processing round, Alice and Bob communicate though an authenticated channel and select only the time-slots when their bases have agreed. It is worth noting that only the time slots corresponding to a measurement result are considered, as in practice neither all pulses sent by Alice contain photons – due to imperfect photon sources – nor all received pulses can be detected by Bob – due to imperfect detectors and channel and component loss.

### B. The Avalanche Photodiode

The traditional single-photon detector used in QKD systems is composed of a cooled InGaAs avalanche photodiode (APD) operated in gated Geiger mode and an avalanche quenching circuit [1,36-38]. When reversely polarized above the breakdown threshold – during short time windows – the device becomes single-photon sensitive, so impinging weak light may trigger a self-sustained avalanche. With the end of the gate, the overvoltage bias is reduced and the macroscopic burst current is quenched to reset the single-photon sensitivity. A discriminator circuit creates a formatted voltage pulse that indicates the occurrence of a photon count. APD-based single-photon detectors are usually not photon number resolving, i.e., they cannot discriminate between a single-photon or multi-photon optical pulse and emit an identical formatted voltage pulse in both cases. Apart the traditional commercial devices, different gating and quenching schemes can also be used with APDs to construct enhanced devices [39,40] and even photon-number resolution capable detectors [41].

### C. Bright light-based attacks: blinding plus faked states

The Geiger mode makes the SPD sensitive to a single photon during the gate time due to the high electronic gain provided by the operational point above the breakdown voltage. However, when biased with lower voltage, the APD works in linear mode, as is usual in telecom applications. In this case, the photocurrent generated in response to optical power is approximately linear. Therefore, if the bias voltage is brought below the breakdown, the device no longer detects single-photons. This behavior can be exploited by Eve to disable Bob's detectors. Of course, this strategy alone gives no information to Eve, but can be combined with other ruses, as discussed ahead.

During this blinding attack, the eavesdropper sends strong light to enforce a current flow, which can alter the threshold breakdown value [15,24-27]. The excess voltage applied to the APD to enable Geiger mode is no longer sufficient and the detector becomes blind to single-photon detection.

Even when operating with low bias voltage (in the case of free-running operation or outside a detection window in gated mode), the SPD may trigger an avalanche if a sufficiently strong pulse is received [20-21]. Assuming an interception-resend strategy, Eve measures each qubit sent by Alice in a random basis and prepares a faked state to send to Bob according to the result obtained. If Bob's basis choice matches Eve's, the strong pulse is routed to the corresponding detector, forcing an avalanche with unity probability, and Eve knows that their results are correlated. However, if their bases disagree, the pulse is split and half the power is delivered to each of Bob's detectors. As this half-power pulse is not sufficient to trigger an avalanche, no detection event takes place. The ruse renders all valid results obtained by Bob correlated to Eve's, who then acquires high knowledge of the key. This strategy may be used standalone (e.g. aftergate attack [17]) or combined to the blinding scheme, causing no critical quantum bit error rate (QBER) increase [20-21], as Bob's SPDs never click in the absence of a faked state. Not so-bright light-based attacks have also been reported, exploring the higher probability of a multi-photon pulse to cause an avalanche [18] or by selectively causing a deadtime on a detector [19].

### D. Detection Efficiency Mismatch

Each SPD operating in gated mode has its own time- (or wavelength-) dependent efficiency curve. This means that the device is usually more efficient in one temporal position (wavelength) relative to the other in a system, due to asymmetries or to temporal mismatch (different responsivity). Eve can explore a mismatch between the two efficiency curves to launch an attack by manipulating, for example, the time of arrival of the qubits relative to the gate windows. A faked states strategy may be employed [10,17,18,20,21], which makes the detection events of Bob more probable to occur at a certain detector according to the delay imposed by Eve. Another example is the time-shift attack [12,13], in which there is no interception at all by Eve, but only a bi-stable random (but known by Eve) delay change in the qubit time-of-flight. By manipulating the optical path, Eve can position the optical pulse in a region of the gate window that increases the probability of a detection occurring in one or other SPD, allowing the eavesdropper to infer (part of) the key without even increasing the QBER. The drawback of this strategy is the reduction of the net detection efficiency of Bob's apparatus that must be compensated by Eve. It is assumed, however, that Eve can replace the link by a more transparent one, or can teleport the states from Alice's output to Bob's entrance.

## III. PROPOSED SCHEME

Here we propose a countermeasure scheme aiming on closing the back-door opened by the SPD regarding its susceptibility to external manipulation by bright light. The spatial modes randomization is presented below. We also report on the experimental implementation of the detector-scrambling strategy, previously suggested in [12], showing that the information of Eve can be hashed out toward zero when random rotation is applied to Bob's detection bases, avoiding the efficiency mismatch-based attacks.

### A. Spatial Modes Randomization

A practical scheme using beamsplitters and additional single-photon detectors is proposed to avoid the direct control of the detectors by the eavesdropper. Figure 1b shows a sketch of Bob's apparatus employed in a BB84-based QKD system with the countermeasure implemented (in contrast, Fig. 1a depicts the traditional BB84 setup). For each received qubit, Bob chooses the measurement basis at his HWP and sends it to a PBS. Each output spatial mode of the PBS is randomly divided in two modes by each beamsplitter (BS) according to

$$|1,0\rangle_{in1,in2} \rightarrow \sqrt{\tau}|1,0\rangle_{out1,out2} + j\sqrt{1-\tau}|0,1\rangle_{out1,out2} \qquad (1)$$

where the indices *in* and *out* represent the two input and two output modes of the BS. When the splitting ratio $\tau$ is 0.5, the photon has the same probability of emerging at one or other output mode. Each output spatial mode is delivered to a SPD.

When under regular operation, whenever Bob's basis is correctly chosen, detector A or B may click for a certain state (say, horizontal SOP); or detector C or D may click if the corresponding orthogonal state is received (vertical SOP). Due to the low average number of photons per optical pulse sent by Alice ($\mu$), and the channel attenuation, the probability of both A and B, or C and D, detectors clicking together (coincident counts) is low. This is given by Poisson distribution as (disregarding the attenuation of the transmission channel and the losses inside Bob's apparatus, for simplicity) $P_{AB} = P_{CD} = [1 - exp(-\mu\eta\tau)]^2$, where $\eta$ is the detection efficiency of the SPDs (assumed to be equal). If $\tau$ is 0.5, both detectors must exhibit similar photon counting statistics and low probability correlated events. When there is no basis agreement, however, any one of the four SPDs clicks at random, as shown in Table 1.

TABLE I
Spatial Modes Occupied by a Single-Photon Pulse in the Regular Setup of the BB84 Protocol and with the Proposed Countermeasure with τ=0.5.

| Alice | Bob's HWP | Regular (Fig. 1a) | Spatial modes countermeasure (Fig. 1b) | Bit |
|---|---|---|---|---|
| |H⟩ | 0° | |A⟩ | (|A⟩+|B⟩) / √2 | 0 |
| | 45° | |C⟩ | (|C⟩+|D⟩) / √2 | 0 |
| | 22.5/67.5° | (|A⟩+|C⟩)/√2 | (|A⟩+|B⟩+|C⟩+|D⟩) / 2 | - |
| |V⟩ | 0 | |C⟩ | (|C⟩+|D⟩) / √2 | 1 |
| | 45 | |A⟩ | (|A⟩+|B⟩) / √2 | 1 |
| | 22.5/67.5° | (|A⟩+|C⟩)/√2 | (|A⟩+|B⟩+|C⟩+|D⟩) / 2 | - |
| |D⟩ | 0 /45° | (|A⟩+|C⟩)/√2 | (|A⟩+|B⟩+|C⟩+|D⟩) / 2 | - |
| | 22.5° | |C⟩ | (|C⟩+|D⟩) / √2 | 0 |
| | 67.5° | |A⟩ | (|A⟩+|B⟩) / √2 | 0 |
| |A⟩ | 0 /45° | (|A⟩+|C⟩)/√2 | (|A⟩+|B⟩+|C⟩+|D⟩) / 2 | - |
| | 22.5° | |A⟩ | (|A⟩+|B⟩) / √2 | 1 |
| | 67.5° | |C⟩ | (|C⟩+|D⟩) / √2 | 1 |

The proposed counter-measure aims on avoiding the control of the SPDs by the eavesdropper, what is accomplished through the randomization of the spatial modes by the two beamsplitters. It is thus essential for the scheme to work that the splitting ratio of the BS cannot be manipulated by the eavesdropper. It is known, however, that the splitting ratio of these devices is intrinsically wavelength dependent [42], and also the isolator's insertion loss [43]. This means that a narrow filter must be inserted at Bob's input to protect the BSs, as a Bragg grating in reflective mode with a circulator, for example. The free spectral range of a Bragg grating is sufficiently long to provide protection over the main spectral range in which the InGaAs-based SPD has significant detection efficiency (the Bragg resonance at the doubled frequency lies in the 775 nm region, considering the operation around 1550 nm). Furthermore, narrow filtering is a key element for the case of coexistence between classical and quantum communication over the same optical fiber links [44].

Consider that Eve launches bright optical pulses (in the bright-light attack context) to deterministically enforce an avalanche on Bob's right detector(s) whenever their bases agree; or cause no avalanche if their bases are mismatched. This behavior is obtained due to power splitting at the PBS according to the SOP after basis choice, and a threshold power level can be defined related to this binary behavior. For simplicity (without lack of generality), suppose that Eve launches a H-polarized pulse with power $P_0$. If Bob chooses rectilinear basis – which occurs with fifty-percent probability – full power emerges from the PBS to arm AB and the power splits to detectors A and B (see Fig. 1b) according to the BS split ratio as $P_A/P_0 = \tau$ and $P_B/P_0 = 1 - \tau$.

If the bases mismatch, half the input power emerges from each output mode of the PBS and, after splitting on the BSs, reach each detector as $P_A/P_0 = P_C/P_0 = \tau$ and $P_B/P_0 = P_D/P_0 = 1 - \tau$.

When under attack, neither the blinding mechanism nor the faked states can operate selectively. If $\tau$=0.5, Eve can manage to set an attack point that is valid on both detectors, she is able to enforce coincident avalanches on both detector ever – this is the symmetric case. This is the best scene from Bob's point of view, as him can easily verify an abnormal amount of coincidences between A and B, or C and D, highly above the Poisson expectation value, what will smoke the eavesdropper out, indicating the presence of the strong pulse.

A more pessimistic case can be analyzed, when, in the matched bases case, Eve can enforce an avalanche on one detector (A) and none on the other on the same arm (B) – the asymmetric case. This can occur depending on the (fixed) asymmetry of the BSs ($\tau \neq$0.5). For example, Eve can send pulses capable of triggering an avalanche on detector A for sure and no avalanche at all on detector B; or can send a stronger pulse which fires both devices simultaneously (taking care not to exceed the threshold power when the bases disagree). Nevertheless, the counting distribution of detectors A and B on both symmetric and asymmetric cases diverges from the expectation of Bob and Eve's ruse still can be detected.

In this context, the use of four standard SPDs behind the two BSs passively creates random spatial modes that are not deterministically accessible to the eavesdropper. The blind strategy suffers from the same limitations, i.e. the blinded detector will always be the same, or both detectors will be always blinded, when the attack is performed. This works as a watchdog against the attacks and has advantages over a monitoring tap placed at the entrance of Bob's station working as an auxiliary watchdog, since it causes no reduction on the system efficiency – the tapped portion of light is not effective for QKD. Furthermore, as discussed in [15,20], there is no guarantee that a tap-based watch-dog with classical or quantum detector cannot be manipulated by Eve, or if there is clear set point to warn Bob about the hacking. The drawbacks of our countermeasure are the increase in the resources employed and a rise in the dark count rate by a factor of 2.

Asymmetry of the BS is not the critical point, but a possible asymmetry inversion caused by Eve, as discussed before. If Eve can deterministically manipulate the value of $\tau$ the security of the QKD section can be fully jeopardized. For example, suppose that Eve is able to set $\tau(\lambda_1)$=0.7 and $\tau(\lambda_2)$=0.3 and that the threshold power level is equivalent to a power loss of 0.5 relative to Bob's input. So Eve can mimics the equiprobable occurrence of events at detectors A and B by sending pulses at $\lambda_1$ (unitary probability for click at SPD$_A$) or at $\lambda_2$ (unitary probability for click at SPD$_B$) by the same amount.

Provided Eve cannot manipulate the splitting ratio of the BSs, let us focus on the case of an asymmetric BS. How well could Eve mimic the expected count ratios between the detectors, or redefining the question, how well can Eve mask the lack of counts enforced on the SPD located at the less transmissive output port of the BS?

A first approximation considers that Eve intercepts Alice's pulses and attack Bob's detectors with bright light for a fraction $\gamma$ of the pulses, while employs a "regular" intercept-resend strategy for the remaining fraction $1-\gamma$ of the pulses. Consider that the bright-light attack causes a binary response of the detectors. Again we concentrate in the case when Alice sends H-polarized states. Mapping the probabilities in a causality tree, as in [42], we get the results for each possibility of the independent basis choice from Eve and Bob. For the sifted events – the selected cases when Alice's and Bob's bases agree – it is found that the bright-light attack causes no error, but introduces (full) asymmetry between detection on $SPD_A$ and $SPD_B$; while the intercept-resend attack introduces error (up to 25%), but causes no asymmetry between de count rates of that detectors. The bright-light attack also reduces the overall detection rate by a factor 2, due to the fact that if the bases of Eve and Bob disagree (half the cases), no event occurs. In this context, the sifting error introduced by Eve is given by $E = 1/4(1-\gamma)/(1-\gamma/2)$, which is lower bounded at zero, if only bright-light attack is performed ($\gamma=1$); and upper bounded at 0.25, if only intercept-resend attack is employed ($\gamma=0$).

The ratio between the detection rates at $SPD_A$ and $SPD_B$ is given by

$$r = (1-\gamma)/(1+\gamma/3) \qquad (2)$$

This means that full bright-light attack results in total assymetry between counts on both detectors, while full intercept-resend attack causes no assymetry at all.

A second approach considers that Eve intercepts Alice's pulses and attack Bob's detectors with bright light only for a fraction $\gamma$ of the pulses, while the remaining fraction $1-\gamma$ is only bypassed to Bob. This strategy belongs to the most general class of attacks [45]. The main advantage of this strategy is that Eve causes no sifting error, independent of the value of $\gamma$. On the other hand, the ratio between the detection rates follows eq. (2), exactly as in the previous case, so the ruse is preferable by Eve's point of view.

Equation (2) gives the big picture for the degree of protection provided by our proposed counter-measured against bright-light-based attacks when fixed asymmetry of the BSs is considered. The fraction of events stolen by Eve is directly related to the asymmetry caused by the attack and the protection is directly given by the ability of Bob to check the asymmetry out. For example, if Bob is able to estimate the detection rate ratio with 1% of uncertainty, he can assume that Eve may have stolen 0.7% of information.

### B. Detector Scrambling

The creation of the proposed random spatial modes does not directly avoid the detection-efficiency mismatch-based attacks (but increases the complexity level required for Eve's ruse). The randomization of the detectors used for photon counting at Bob's station is, nonetheless, sufficient to close this backdoor, while no additional hardware is needed.

For each incoming qubit, Bob not only randomly chooses the detection basis, but also may randomly invert the attribution of the pairs of detectors. This means that, if Bob wants to use the rectilinear basis, the HWP may be turned to 0º or 45º (or to 22.5º or 67.5º, if the diagonal basis is chosen). The additional choice [46] has no impact on the final value of the bit shared between Alice and Bob. For example, whenever Alice sends a H-polarized qubit, the bit inferred by Bob with matched basis is deterministic, say "0", but the click can occur at either detector A' (A or B in Fig. 1), if the HWP is set to 0º; or at detector C' (C or D in Fig. 1), if the HWP is set to 45º. The same idea holds for the diagonal basis. The process is written as the randomization of Alice's SOP to the spatial modes A and C as:

$$|H\rangle_{Alice} \rightarrow \cos 2\theta \; |1,0\rangle_{A,C} + \sin 2\theta \; |0,1\rangle_{A,C}$$
$$|V\rangle_{Alice} \rightarrow \sin 2\theta \; |1,0\rangle_{A,C} + \cos 2\theta \; |0,1\rangle_{A,C}$$
$$(3)$$

where $\theta$ is the physical angle of Bob's HWP.

Provided that this choice is truly random, Eve cannot infer at which SPD Bob's detection has occurred, so the final spatial mode is not accessible to her. In the time-shift attack context, if Eve waits the basis reconciliation and learns that Alice and Bob agreed, there is no way to infer which version of the basis Bob has chosen and the detector used, based on the imposed delay, even if the efficiency curves are fully mismatched. The countermeasure is equivalent to an active randomization of the spatial modes and drastically reduces the mutual information between Alice-Eve and Bob-Eve. When Bob's basis matches Alice's, the logical result is deterministic, but the detector that registered the event is random.

There is some discussion [13] if a Trojan-horse attack [47] could be attempted against the scrambler (the HWP in our case). Eve could send a strong pulse and read out the polarization rotation by analyzing the measured backscattered light. As discussed before, narrow filtering and isolators are necessary at the input of Bob's apparatus to provide technological protection. Here the spatial modes randomization acts in the same way as described in Section III.A to avoid the Trojan-horse attack in a fundamental level. As the probe pulse used in the attack must be bright – to overcome the isolation from Bob's HWP back to Eve – the photon counting statistics of Bob's SPDs will behave differently as from the expected.

### IV. EXPERIMENTAL SETUP

#### A. Spatial Modes Randomization

The countermeasure against quantum hacking based on randomized spatial modes was experimentally implemented, as depicted in Fig. 2, which emulates the aftergate attack [17].

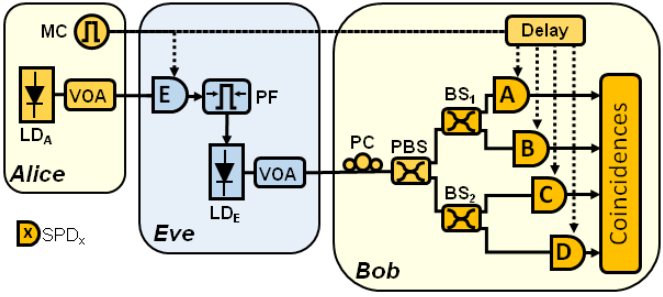

Fig. 2. Experimental setup for the passive randomization of the spatial modes as a countermeasure for attacks using faked states. The setup emulates the aftergate attack. LD: laser diode; VOA: variable optical attenuator; MC: master clock; PF: pulse formatter; PC: polarization controller; PBS: polarizing beamsplitter. BS: beamsplitter; A-E: single-photon APDs.

During the attack, Eve intercepts the optical pulses sent by Alice and resend faked states to Bob according to her result. The statistics of the intercepted photons is obtained in the experiment by the detection of a continuous-wave (CW) faint laser diode source (LD) sent by Alice though a variable optical attenuator (VOA). Eve detects the photons with a single-photon APD (labeled E) operating in gated-Geiger mode. The detector has 15% detection efficiency and opens 2.5-ns wide gates triggered by the system's master clock (MC) at 100 kHz. A 10 μs deadtime is enforced after each detection event to avoid the afterpulse effect. Whenever a detection event occurs at Eve's SPD, the voltage pulse created by the detector is compressed by a pulse formatter (PF) and drives Eve's laser diode source (LD$_E$). The bright-light faked states are sent by Eve to Bob's station. The optical power of the pulses is regulated by a VOA and are delayed to reach Bob's detectors at the attack point, at the end of the gate (hence aftergate attack), as explained ahead.

Bob's station setup is similar to the concept depicted in Fig. 1b, and is fully composed by fiber-optical elements. A polarization controller (PC) acts as a HWP for the basis selection. Actually, this SOP selection is shared by Eve and Bob in this setup, and their combined angles are, in fact, emulated. A PBS performs the state projection and two BS create the passive randomized spatial modes, where the four single-photon APDs are placed (labeled A to D). The SPDs are similar to Eve's, including their configuration. An FPGA-based coincidence counting module is connected to the four SPDs to acquire the single and coincident detection events. The delay generator is used to trigger the four SPDs in a way that the strong pulse reaches them at the end of the gate, when under the aftergate attack; or inside the gate, when under regular (no-attack) operation. We note that in a real attack, Eve herself would manipulate the relative delay between the faked states and the SPDs.

To find the attack set point, a 1-ns wide optical pulse was scanned though the detection gate of each one of the four SPDs used by Bob in the experiment. As the pulse peak power is increased, the gate end is extended, due to the residual overvoltage bias. The set point for the aftergate attack, shown in Fig. 3, is reached when a 3-dB step in peak power is sufficient to allow a binary behavior of the counting probability.

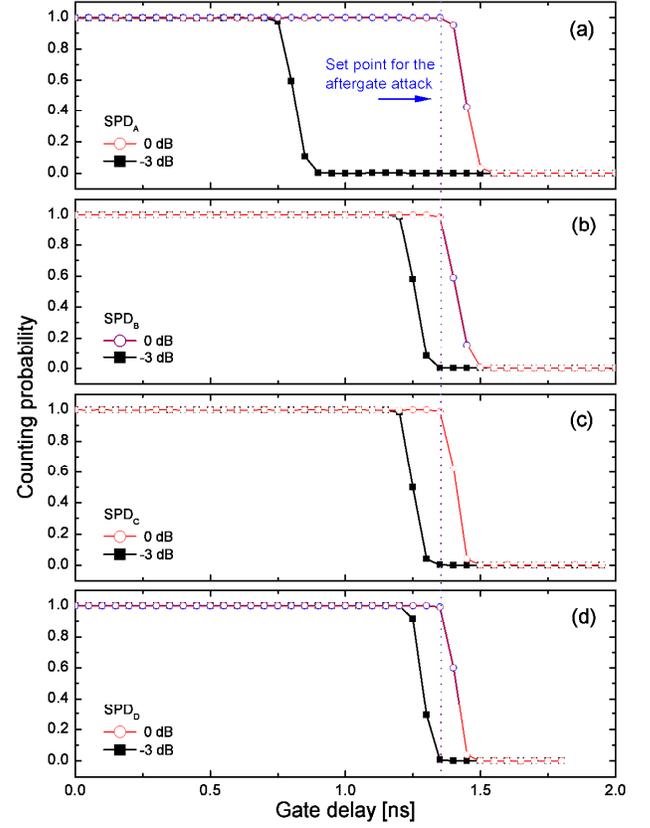

Fig. 3. Scan of optical pulses with 3-dB different peak power though the gate of the four SPDs used in the experiment (each subfigure corresponds to one of Bob's SPDs). The set point for the aftergate attack, when a binary response of the device is obtained, is indicated. All detectors are of the same model, except for SPD$_A$. The higher separation of the curves for that detector is probably due to a different internal gate waveform.

The time delay between the bright pulses and the detection gates is then fixed as indicated in Fig. 3. The 3-dB step is equivalent to the cases when Bob's and Eve's bases match (higher power reaching the right SPDs) or not (half-power to each detector).

In the experiment, Eve sends horizontal or vertical SOPs to Bob, who sets the rectilinear or diagonal measurement bases with a HWP adjusted to 0° or 22.5°, respectively corresponding to matched and mismatched bases relative to Eve. The single and coincident events are collected during 300 s for both cases with regular operation and under attack.

## B. Detector Scrambling

The detector-scrambling countermeasure was also experimentally implemented, as depicted in Fig. 4, which emulates the time-shift attack [12].

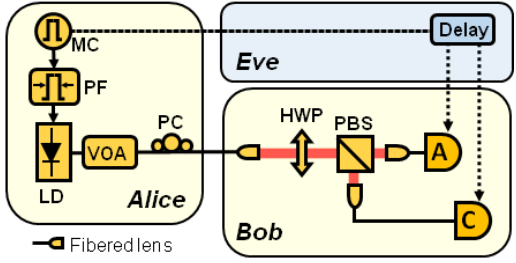

Fig. 4. Experimental setup for the detector-scrambling countermeasure against detection efficiency mismatch-based attacks. This setup emulates the time-shift attack. LD: laser diode; PF: pulse formatter; MC: master clock; VOA: variable optical attenuator; PC: polarization controller; HWP: half-wave plate; PBS:polarizing beamsplitter; A,C: single-photon APDs.

Alice sends 1-ns wide faint optical pulses through a variable optical attenuator (VOA) with horizontally- or vertically-oriented SOPs to Bob. Alice's laser diode source (LD) is driven by the formatted pulses from the master clock (MC). Bob launches the incoming pulses in free-space through a collimating fiber-pigtailed lens. The beam passes through a bulk HWP and a bulk PBS. The output modes of the PBS are collected with aspheric lenses and delivered to SPDs, set as in the previous experiment. Two SPDs were used for this proof-of-principle of the detector-scrambling strategy. The HWP can be set to 0° or 45° – the two versions of the ⊕ basis. The bulk elements make the adjustments easier in the experiment, but in practice a fiber-optical modulator must be employed. In QKD systems using active basis choice, which correspond to most commercial systems, the randomizing element can be the same one that Bob uses for his basis choice. The only difference is that a set of four different rotations become necessary, instead of two, which requires the use of two random numbers per clock.

In the time-shift attack, Eve controls the time-of-flight of the pulses between Alice and Bob. Here, we analogously emulate the time shift by acting on the relative electronic delay between the optical pulses and the detection gates, all driven/triggered by the master clock. The relative gate delay of Bob's detectors was scanned relative to the optical pulse and their normalized detection efficiency was measured for both devices.

## V. RESULTS

The single and coincident counts were acquired for the system described in Fig. 2 with regular operation and under the aftergate attack. Both results are show in Fig. 5a and 5b, respectively, acquired with the HWP set to different angles.

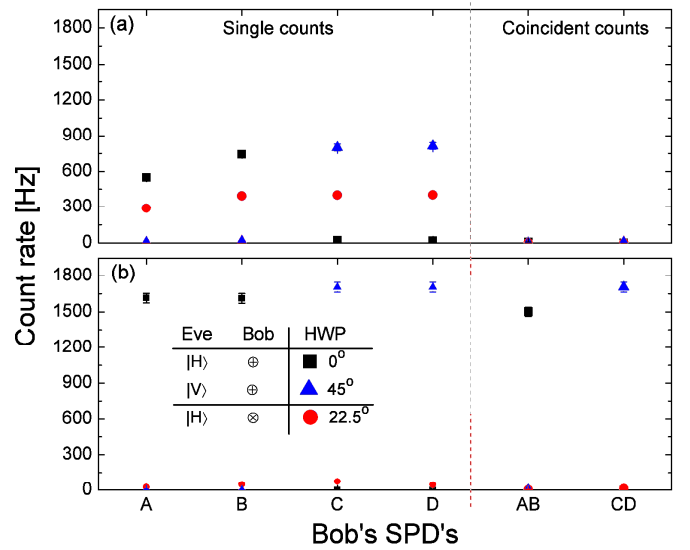

Fig. 5. Count rate of the four SPDs (a) with regular operation and (b) under the aftergate attack. The HWP is shared by Alice and Bob in the experiment, so the set angles represent the composition of their devices.

With regular operation, when the HWP is set to 0°, Eve sends horizontally-oriented SOP and Bob measures in the matching rectilinear basis, so $SPD_A$ or $SPD_B$ click (excepting for dark counts and residual optical misalignment effects). When the HWP is set to 45°, the pulses sent with vertical SOP are measured in the matching rectilinear basis. Note that the HWP is shared by Eve and Bob in the experiment, so the set angle corresponds to the composition of their individual devices. The pulses are then routed to the other output mode of the PBS, and $SPD_C$ or $SPD_D$ click. When the wrong basis is chosen for measurement, with HWP set to 22.5°, there is not a preferential branch to click and any detector may fire.

Note the comparative behavior of the single counts with matched bases: when under regular operation, each faint pulse is randomly routed to one of the two detectors of the corresponding branch after the PBS; on the other hand, when under attack, a 3-dB level appears in the count rate of the two detectors, as both of them simultaneously fire due to the bright optical pulse. This ratio must be hidden by Eve, which can be accomplished by sacrificing half the detections at the interception step.

The key point of the countermeasure appears when analyzing the coincident counts when the symmetric-BS case is assumed. Under normal operation, a small fraction of coincident counts is expected, between $SPD_A$ and $SPD_B$ or between $SPD_C$ and $SPD_D$, due to multi-photon pulses that are split in the BSs and dark counts. As seen in Fig. 5a, this occurrence is very low. On the other hand, when under the aftergate attack, almost all detections (with matched bases) are coincident, i.e., Eve always enforces detections in both $SPD_A$ and $SPD_B$ or in both $SPD_C$ and $SPD_D$, simultaneously. Eve's fingerprints are clear from the results of the coincident counts. The coincident events between the other pairs of SPDs are null. Here the symmetrical hypothesis of Section III.A is assumed, but the (fixed-) asymmetrical hypothesis will also leave a strong signature, as only one detector of each pair will usually click – the fixed-asymmetric-BS case.

Another fingerprint left by Eve is observed when the bases are not matched. Although these events are discarded by Alice and Bob in the sifting procedure, this is avaluable source of information to Bob regarding an external intervening, as the occurrence of uncorrelated photon counts is severely reduced when the attack is performed, as seen in Fig. 5.

The results for the time-shift attack and the detector-scrambling countermeasure, measured with the setup of Fig. 4, are shown in Fig. 6.

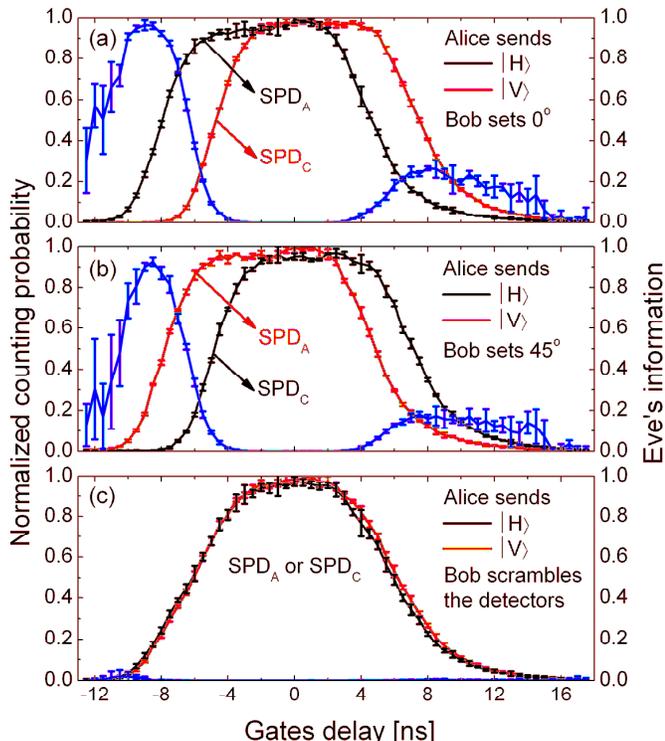

Fig. 6. Normalized counting probability at Bob's detectors $SPD_A$ and $SPD_C$ when Alice sends H- (black lines) or V-polarized (red lines) states and Bob uses $\oplus$ basis for measurement. (a) Traditional operation (Bob's HWP set to 0°); (b) flipped detectors (Bob's HWP set to 45°); (c) averaged value with scrambled detectors (Bob's HWP is randomly flipped between 0° and 45° for each generated pulse). The mutual information between Eve and Bob depends on the ratio between the counts at the detectors and appears as the blue lines in (a)-(c). The error bars come from the standard deviation of the counts.

Figure 6 shows the case of temporal efficiency mismatching between the gates of $SPD_A$ and $SPD_C$. The normalized counting probability at $SPD_A$ and $SPD_C$ are shown when Alice respectively sends H- (black curves) or V-polarized (red curves) weak coherent states and Bob uses the $\oplus$ basis for measurement. For each gate delay, the mutual information between Eve and Bob is calculated from the ratio of the curves of both detectors as [12]

$$I_E = 1 - H_2\left(\frac{r}{1+r}\right) \qquad (4)$$

where $r$ is the minimum ratio of the efficiency curves at a given delay and $H_2(x)$ is the Shannon entropy. The mutual information between Eve and Bob is also displayed the figure (blue lines).

The results in Fig. 6a are obtained with the scan of the optical pulse through the detection gates when Bob's HWP is set to 0° ($\oplus$ basis), resulting in higher count rates in $SPD_A$ and $SPD_C$, respectively, for H- and V-polarized states sent by Alice. This is the conventional operation without the detector-scrambling countermeasure. Eve's information can reach unity depending on the attack point (referring to the pulse delay enforced by Eve, i.e., the position of the optical pulse in the gate delay axis). If Eve enforces a set point to exploit the detection efficiency mismatch, her information about the resulting detection increases due to the higher probability of the event having occurred in the more efficient SPD.

Fig. 6b exhibits equivalent results: Bob sets the HWP to 45° ($\oplus$ basis with flipped SPDs), causing the detectors assignment to be inverted, therefore $SPD_A$ and $SPD_C$ now detect vertical and horizontal SOPs, respectively. The mutual information of Bob and Eve stills reaches high levels when looking only at this case.

When the countermeasure is active in a QKD session, Fig. 6a and 6b randomly occur, resulting in the averaged detection probability shown in Fig. 6c. We see that the efficiency mismatching disappears, and both curves corresponding to the detection of H- and V-polarized states are fairly identical. The causes of such an asymmetry are mainly the finite optical misalignment error of the SOPs and of the HWP (around 0.5%) and the statistical dispersion of the data. The first cause can be improved by careful polarization alignment of the system, while the second cause can be reduced by improving the statistics of the collection of counting events.

The mutual information between Eve and Bob drops significantly towards zero (the ideal value). A residual amount of information (due to the asymmetry discussed before) can be seen in Fig. 6c, but we can also see an efficiency penalty that must be compensated by Eve to cover her fingerprints: since the detectors are scrambled, Bob's apparent efficiency is actually the average of the mismatched counting probabilities. If Eve chooses to delay the photons to a relative temporal position with allows for higher information leakage, but far from the efficiency peak (as in the delay -10 ns on Fig. 6c), the efficiency perceived by Bob will be considerably lower than the original expected peak values. The information an eavesdropper can extract, calculated from the normalized values, is reduced from the initial value of 97%, shown in Fig. 6a, to values below 2% with the countermeasure active, shown in Fig. 6c.

## VI. CONCLUSION

Despite the fact that a full practical solution for all kinds of quantum hacking attacks aimed at all aspects of traditional QKD systems has not yet been found or proved to be possible, the recent overflow of eavesdropping schemes motivated the proposal of many practical solutions. We have shown how some back-doors at the detection end can be closed in a standard BB84 frame with the creation of randomized spatial modes, passively by a combination of beamsplitters and extra SPDs and by actively scrambling the detectors at the measurement station. This represents a practical and readily implementable solution against bright-light- and efficiency-mismatching-based quantum hacking attacks aimed on the single-photon detector so far.